\documentclass[12pt]{article}
\usepackage{amsmath}
\usepackage{amssymb}
\usepackage{amsfonts}
\usepackage[colorlinks=true,linkcolor=RawSienna,citecolor=RawSienna,urlcolor=RawSienna,bookmarksopen=true,pdfstartview=FitB]{hyperref}
\usepackage[dvipsnames]{xcolor}
\usepackage[top=1.1in, bottom=1.1in, left=1in, right=1in]{geometry}
\usepackage[onehalfspacing]{setspace}
\usepackage{natbib}
\usepackage{algorithm}
\usepackage{algpseudocode}
\usepackage{graphicx}
\usepackage{caption}
\usepackage{subcaption}
\usepackage{enumerate}

\let\Algorithm\algorithm
\renewcommand\algorithm[1][]{\Algorithm[#1]\setstretch{1.15}}
\newtheorem{theorem}{Theorem}[section]

\newtheorem{lemma}{Lemma}[section]

\begin{document}

\title{\vspace{-1.2cm} Estimating The Proportion of Signal Variables Under\\ Arbitrary Covariance Dependence}
\author{X. Jessie Jeng\thanks{%
Address for correspondence: Department of Statistics, North Carolina State University, 2311 Stinson Dr., Raleigh, NC 27695-8203, USA.  Email: \texttt{xjjeng@ncsu.edu}.}\\
Department of Statistics, North Carolina State University }
\date{}
\maketitle

\begin{abstract}

Estimating the proportion of signals hidden in  a large amount of noise variables is of interest in many scientific inquires. In this paper, we consider realistic but theoretically challenging settings with arbitrary covariance dependence between variables. We define mean absolute correlation (MAC) to measure the overall dependence level and investigate a family of estimators for their performances in the full range of MAC.  We explicit the joint effect of MAC dependence and signal sparsity on the performances of the family of estimators and discover that 
no single estimator in the family is most powerful under different MAC dependence levels. Informed by the theoretical insight, we propose a new estimator to better adapt to arbitrary covariance dependence. The proposed method compares favorably to several existing methods in extensive finite-sample settings with strong to weak covariance dependence and real dependence structures from genetic association studies.

\medskip

   
\end{abstract}


\section{Introduction}

We consider the problem of estimating the proportion of information bearing signals that are sparsely located in a large amount of noise variables. This problem is of interest in many scientific inquiries. For example, estimation of signal proportion is required by multiple testing methods to calculate local false discovery rate \citep{efron2007size}, to derive q-value \citep{storey2003positive}, and to improve power \citep{storey2002direct, finner2009controlling}. 
Moreover, in many multi-stage studies, estimation of signal proportion can assist efficient pre-screening and sample size calculation \citep{cai2017optimal}. 
A recent line of research, which focuses on retaining a high proportion of signals through efficient false negative control, also replies on the estimation of signal proportion as a benchmark for signal inclusion \citep{jeng2016rare, jeng2019efficient}. 

Although estimation of signal proportion is widely requested, methodology development has met two major challenges. First, signals of different sparsity levels often call for different estimation methods, while signal sparsity levels are unknown a priori. Secondly, the large set of variables under investigation may have complex dependence structures. 
There exist a number of rigorously developed methods. 
Most of them, however, assume independence between variables \citep{GW04, MR06, jin2007estimating} and sparsity levels within a certain range \citep{cai2007estimation, jin2008proportion}. An extensive review of the existing methods can be found in \cite{chen2019uniformly}.
More recent developments extend the study to consider specific dependence structures. For example, \cite{jeng2019efficient} studies the problem assuming block-diagonal covariance structures for the variables; \cite{jeng2019variable} considers the problem in linear regression and imposes certain dependence and sparsity conditions to facilitate accurate precision matrix estimation and bias mitigation.  There lacks a method to consistently estimate signal proportion under arbitrary covariance dependence when signal sparsity is unknown and possibly falls in a wide range. Such an universal estimator can have far-reaching impact in real applications. 

In this paper, we define mean absolute correlation (MAC) to measure the overall covariance dependence level and investigate a family of estimators for their performances in the full extent of MAC. 
We explicate the joint effect of MAC dependence and signal sparsity on the performances of the family of estimators and discover that the most powerful estimator under independence is no longer the best candidate under arbitrary covariance dependence. Moreover, it shows that no single estimator in the family is most powerful under different MAC dependence levels. 
Informed by the theoretical insight, we propose a new estimator to better adapt to arbitrary covariance dependence and signal sparsity.


The new method is compared with several popular methods in extensive simulation examples including strong to weak covariance dependence between variables and real dependence structures from genetic association studies. 
It shows that although the winner of the several existing methods changes over different settings, performance of the new method is either comparable to or better than the performance of that winner in each setting. 
We apply the new method to analyze two real datasets.  The first dataset is from an expression quantitative trait loci (eQTL) study with 8637 candidate single-nucleotide polymorphisms (SNPs), for which the overall dependence in terms of the MAC level is rather weak.   
The second dataset is from a classical association study with microarray data, where 4088 candidate genes possess much stronger overall dependence with a high MAC level.
We compare the estimates of the new method with those of the existing methods and 
validate the results through simulations under the same dependence structures. Our method exhibits better accuracy for signals of different signal sparsity levels under the real dependence structures. 

The rest of the paper is organized as follows. Section \ref{sec:method} first introduces a family of lower bound estimators and develops a general result on their estimation consistency under arbitrary dependence. Then, for specific estimators in the family, the joint effects of signal sparsity and covariance dependence are explicated. Consequently, a new and more powerful estimator is developed under arbitrary covariance dependence.  
Section \ref{sec:simulation} compares the proposed method with existing methods in simulation examples. Section \ref{sec:application} applies the new and existing methods to real genetic association studies. Section \ref{sec:conclusion} concludes the work with further discussions. Technical proofs are provided in Appendix.

\section{Method and Theory} \label{sec:method}

Denote $I_0$ and $I_1$ as the sets of indices for signal and noise variables, respectively. We consider the marginal distribution of $p$ variables as    
\[
X_j \sim F_{0} \cdot 1\{j \in I_0\} + F_1 \cdot 1\{j \in I_1\}, \qquad j=1, \ldots, p, 
\]
where $F_0$ and $F_1$ are the null and signal distribution, respectively. Define the signal proportion 
\[
\pi = |I_1|/p.
\] 
We assume that $F_0$ is continuous and known a priori. All the other components are unknown. Our goal is to estimate the signal proportion $\pi$ without the need to specify $F_1$ or to identify which variables are signals.

\subsection{A family of estimators} \label{sec:family}

\cite{MR06} introduced a family of proportion estimators that are built upon the empirical process of $p$-values under independence. This family of estimators have been proved to provide lower bound estimates for the true proportion $\pi$. Members in the family are indexed by the choice of a bounding function, and it has been shown that the estimator with 
bounding function $\delta(u) = \sqrt{u(1-u)}$ has the best overall performance for signals of different sparsity levels. This conclusion, unfortunately, does not hold anymore under arbitrary dependence. 
When variables are arbitrarily dependent, the limiting distribution of the empirical process of $p$-values is generally unknown and may not even have an analytic expression. Moreover, the dependence effect mingles with signal sparsity to influence the performances of different estimators in the family. These difficulties substantially complicate the estimation problem and motivate us to develop new techniques to study the family of estimators and to come up with a new powerful estimator under arbitrary covariance dependence.   

For presentation simplicity, we perform inverse normal transformation as $Z_j = \Phi^{-1}(F_0(X_j))$, where $\Phi^{-1}$ is the inverse of the cumulative distribution function of a standard Normal distribution. Then, we have
\begin{equation} \label{def:modelZ}
Z_j \sim \Phi \cdot 1\{j \in I_0\} + G \cdot 1\{j \in I_1\}, \qquad j=1, \ldots, p, 
\end{equation}
where $G$ denotes the signal distribution after inverse normal transformation, which remains unknown. 

Next, we construct a modified family of estimators to accommodate dependence among $Z_j$. Let 
\[
\bar W_p(t) = p^{-1}\sum_{j=1}^{p} 1\{|W_{j}|>t\},
\] 
where $(W_1, \ldots, W_p)$ follow the joint null distribution of $(Z_1, \ldots, Z_p)$. 
Denote $\bar{\Phi}(t) = 1-\Phi(t)$. For a given function $\delta(t)$ that is strictly positive on $(0, \infty)$, define
\begin{equation} \label{def:Vp}
V_{p, \delta}=\sup_{t> 0}\frac{|\bar W_p(t)-2\bar{\Phi}(t)|}{\delta\left( t\right) }.
\end{equation}
Apparently, $V_{p, \delta}$ varies with the choice of $\delta(t)$, and $\delta(t)$ is called a bounding function. For a given $\delta (t)$ and a control level $\alpha$, we define the corresponding bounding sequence $c_{p, \delta}$ as a function of $p, \delta$ and $\alpha$ that satisfies the following properties:

(a) $p c_{p, \delta} >p_0 c_{p_0, \delta}$, where $p_0= |I_0|$, and 

(b) $P(V_{p, \delta} > c_{p, \delta}) < \alpha$ for all  $p$. \\
From the above properties, it can be seen that $c_{p, \delta}$ is an upper bound of $V_{p, \delta}$, while $V_{p, \delta}$ relies on the joint null distribution.
Both $V_{p, \delta}$ and $c_{p, \delta}$ carry the information of dependence among the variables. Also, the absolute sign in the numerator of (\ref{def:Vp}) stabilizes  $V_{p, \delta}$ as $\bar W_p(t)-2\bar{\Phi}(t)$ may not be asymptotically symmetric anymore under arbitrary dependence. 

Then, given the observed $Z_j$, a family of estimators indexed by $\delta$ are constructed as
\begin{equation} \label{def:MR}
\hat{\pi}_{\delta}=\sup_{t > 0}{\frac{\bar F_{p}(t)-2\bar{\Phi}(t)-c_{p, \delta}\delta \left( t\right) }{1-2\bar{\Phi}(t)}}, 
\end{equation}
where 
\[
\bar F_{p}(t) = p^{-1} \sum_{j=1}^{p} 1\{|Z_{j}|>t\}.
\] 
It can be seen that the $\hat \pi_\delta$ family of estimators hinge on the choices of the bounding function $\delta(t)$ and the corresponding bounding sequence $c_{p, \delta}$. As the bounding sequence $c_{p, \delta}$ implies a normal range of $\bar F_p(u)$ if all variables are noise, the information carried by the observed $\bar F_p(u)$ that exceeds the normal range presents evidence for the existence of signals. 
This version of lower bound family is for signals of two-sided effects. Minor changes to accommodate one-sided signal effect is straightforward. Details for the numerical implementation of $\hat \pi_\delta$ can be found at the end of Section \ref{sec:estimator}.

As shown in \cite{MR06}, for signals of different sparsity levels, the performances of the estimators in the original family can be very different under independence. 
More specifically, the signal proportion can be re-parameterized as $\pi = p^{-\gamma}$, $\gamma \in (0, 1)$, with $\gamma \in (0, 1/2)$ representing the relatively dense case and $\gamma \in [1/2, 1)$ representing the more sparse case. Consistency of different estimators in the original family was proved under independence for $\gamma \in (0, 1/2)$ and $\gamma \in [1/2, 1)$ separately. Here, we consider the estimation problem in the more challenging setting with arbitrary dependence and the modified family of estimators to accommodate the dependence. 

First, we impose a monotonicity constraint on the bounding function $\delta(t)$, which allows us to present the following result under general dependence for signals of sparsity levels in the full range of $\gamma \in (0,1)$.

\begin{theorem} \label{thm:general_MR}
	Consider model (\ref{def:modelZ}). For a given bounding function $\delta(t)$, if there exist a bounding sequence $c_{p,\delta}$ satisfying the properties in (a) and (b), then 
	\begin{equation}  \label{eq:lower}
	P(\hat \pi_{\delta} < \pi) \ge 1-\alpha.
	\end{equation}
	On the other hand, for $\pi$ satisfying $0 < \pi \ll  1$,  if $\delta(t)$ is non-increasing with respect to $t$, and $G = G_p$ such that $G_p(\tau) \to 0$ or $G_p(-\tau) \to 1$ for some $\tau$ such that $\tau\gg 1$ and $\delta(\tau) \ll \pi / c_{p, \delta}$, then 
	\begin{equation} \label{eq:upper}
	P(\hat \pi_{\delta} > (1-\epsilon) \pi) \to 1
	\end{equation}
	for any constant $\epsilon>0$. 
\end{theorem}

The above theorem says that the lower bound property of $\hat \pi_\delta$ in (\ref{eq:lower}) holds as long as a bounding sequence  $c_{p,\delta}$ satisfying (a) and (b) can be found. On the other hand, the upper bound property of $\hat \pi_\delta$ in (\ref{eq:upper}) holds under certain conditions on the signal distribution $G$, which essentially says that the signal effect, either positive ($G < \Phi$) or negative ($G>\Phi$), is strong enough. When both lower bound and upper bound conditions are satisfied for a given $\delta(t)$ and a degenerating $\alpha$ such that $\alpha= \alpha_p \to 0$, $\hat \pi_\delta$ consistently estimate the true signal proportion, i.e., for any constant $\epsilon>0$,
\[
P((1-\epsilon) \pi \le \hat \pi_{\delta} < \pi ) \to 1.
\]
The above results hold for arbitrarily dependent variables and for sparse signals that are not of a fixed proportion of all the variables. Such sparsity condition covers the full range of $\gamma \in (0,1)$.  

\subsection{Joint effect of MAC dependence and sparsity}  \label{sec:estimation}

As individual estimators in the $\hat \pi_\delta$ family hinge on the choice of the bounding function, we further study their consistency with specific $\delta(t)$ functions. We focus on $\delta(t)$ of the form 
$\delta(t)=[\bar \Phi(t)]^{\theta}$, $\theta \in [0,1]$, so that the corresponding  $\hat \pi_\delta$ are closely related to several existing estimators whose consistency has been studied under independence in literature (please see \cite{MR06} and the references therein).  Valuable insight can be obtained by comparing the estimators' performances under independence and dependence, which helps us construct a new and more powerful estimator under arbitrary covariance dependence. 


In order to explicate the effect of dependence on  $\hat \pi_\delta$, we assume that 
\begin{equation} \label{model:joint_normal}
(Z_1, \ldots, Z_p) \sim N_p(\mu, \Sigma), 
\end{equation}
where $\mu$ is a $p$-dimensional sparse vector with $\mu_j = A\cdot 1\{j \in I_1\}$, $A\ne 0$,  and $\Sigma$ is an arbitrary correlation matrix, i.e., $\Sigma_{ij} = Corr(Z_i, Z_j)$. For presentation simplicity, assume $A>0$.  
We define the Mean Absolute Correlation (MAC) to calibrate the covariance dependence as  
\begin{equation} \label{def:bar_rho}
\bar \rho_{\Sigma} = \sum _{i=1}^p \sum_{j=1}^p |\Sigma_{ij}| / p^2.
\end{equation}
A larger value of $\bar \rho_{\Sigma}$ indicates stronger overall dependence.


Moreover, we employ a discretization technique from \cite{arias2011global} and \cite{jeng2019variable} as follows. 
Define $\mathbb{T} = [1, \sqrt{5 \log p}] \cap \mathbb{N}$ 
and the discretized version of $V_{p, \delta}$ as
\begin{equation} \label{def:Vp*}
V^*_{p, \delta}=\max_{t \in \mathbb{T}}\frac{|\bar W_p(t)-2\bar{\Phi}(t)|}{\delta (t) }.
\end{equation}
Denote $c^*_{p, \delta}$ as the bounding sequence based on $V^*_{p, \delta}$, and define the corresponding proportion estimator as 
\begin{equation} \label{def:pi*}
\hat{\pi}^*_{\delta}=\max_{t \in \mathbb{T}}\frac{\bar F_{p}(t)-2\bar{\Phi}(t)-c^*_{p, \delta} \delta (t) }{1-2\bar{\Phi}(t)}.
\end{equation}

Next, we explicates how the MAC level interacts with signal sparsity and signal intensity to influence the consistency of $\hat{\pi}^*_{\delta}$ with $\delta(t) = [\bar \Phi(t)]^\theta$ and reveals very different results for $\theta \in [0,1/2]$ and $\theta \in (1/2, 1]$.  

\begin{theorem} \label{thm:discrete_0.5}
	Consider model (\ref{model:joint_normal}). Let $\delta(t) = [\bar \Phi(t)]^\theta$ with $\theta \in [0,1/2]$. Then, there exists a bounding sequence 
	\[
	c_{p, \delta}^* = O\left(\sqrt{\bar \rho_\Sigma (\log p)^{\theta+1/2}}\right),
	\] 
	that satisfies properties (a) and (b), and the corresponding estimator $\hat{\pi}^*_{\delta}$  satisfies $ P(\hat \pi^*_{\delta} < \pi) \ge 1-\alpha$. Moreover, for $\pi$ satisfying $0 < \pi \ll  1$,	
	if $A = A_p$ such that $A_p \gg 1$ and 
	\begin{equation} \label{cond:theta_05}
	A_p - \bar \Phi^{-1}\left( {\pi^{1/\theta} \over \bar \rho_{\Sigma}^{1/(2\theta)} (\log p)^{(\theta+1/2)/(2\theta)}} \right) \to \infty,
	\end{equation}
	then $P(\hat \pi^*_{\delta} > (1-\epsilon) \pi) \to 1$ for any constant $\epsilon>0$.
\end{theorem}
The above theorem says that for $\delta(t) = [\bar \Phi(t)]^\theta$ with $\theta \in [0,1/2]$, we can find a bounding sequence $c_{p, \delta}^*$, whose order increases with respect to $\bar \rho_\Sigma, p$ and $\theta$, respectively.   
The second part of the theorem provides a condition on the signal distribution for the consistency of $\hat \pi^*_{\delta}$. This condition is easier to be satisfied for less sparse signals (larger $\pi$) or less dependent variables (smaller $\bar\rho_{\Sigma}$). Note that $\bar \Phi^{-1}(\pi^{1/\theta} / (\bar \rho_{\Sigma}^{1/(2\theta)} (\log p)^{(\theta+1/2)/(2\theta)}))$ is  well-defined only for $\pi < \sqrt{\bar \rho_{\Sigma} (\log p)^{\theta+1/2}}$. In the case $\pi \ge  \sqrt{\bar \rho_{\Sigma} (\log p)^{\theta+1/2}}$, condition on $A_p$ is simply $A_p \gg 1$. 
For the special case with independent variables, $\bar \rho_{\Sigma} = 1/p$ and(\ref{cond:theta_05}) degenerates to  
\[
A_p -  \bar \Phi^{-1}\left(\pi^{1/\theta} p^{1/2\theta} / (\log p)^{(\theta+1/2)/(2\theta)} \right) \to \infty ,
\] 
which agrees with the sufficient and necessary condition for the consistency of $\hat \pi_{\delta}$ under independence for relatively sparse signals. The comparison can be made by adopting the same parameterization as in Theorem 3 of \cite{MR06} with $\pi = p^{-\gamma}$, $\gamma \in [1/2, 1)$, $\nu = \theta$,  and $\kappa=2$.

Results in Theorem \ref{thm:discrete_0.5} explicate how the performance of $\hat{\pi}^*_{\delta}$ with $\delta(t) = [\bar \Phi(t)]^\theta, \theta \in [0, 1/2]$ deteriorates as the MAC dependence gets stronger. These results, however, cannot be extended to $\hat{\pi}^*_{\delta}$ with $\theta \in (1/2, 1]$. For the latter, we present the following theorem.  

\begin{theorem} \label{thm:discrete_1}
	Consider model (\ref{model:joint_normal}). 
	Let $\delta(t) = [\bar \Phi(t)]^\theta$ with $\theta \in (1/2, 1]$. Then, there exists a bounding sequence $c_{p, \delta}^* = O(\sqrt{\log p})$, that satisfies the properties (a) and (b), and the corresponding estimator $\hat{\pi}^*_{\delta}$ satisfies $ P(\hat \pi^*_{\delta} < \pi) \ge 1-\alpha$. Moreover, for $\pi$ satisfying $0 < \pi \ll  1$,  if $A = A_p$ such that $A_p \gg 1$ and 
	 \begin{equation} \label{cond:theta_1}
	 A_p - \bar \Phi^{-1}\left({\pi^{1/\theta} \over (\log p)^{1/(2\theta)}}\right) \to \infty.
	 \end{equation}
	 then $P(\hat \pi^*_{\delta} > (1-\epsilon) \pi) \to 1$ for any constant $\epsilon>0$.	
\end{theorem}
This theorem shows that for $\delta(t) = [\bar \Phi(t)]^\theta$ with $\theta \in (1/2, 1]$, we can find a bounding sequence $c_{p, \delta}^*$, whose order does not involve the MAC dependence level $\bar \rho_\Sigma$. 
Moreover, the intensity condition in (\ref{cond:theta_1}) does not involve $\bar \rho_{\Sigma}$, which means that $\hat{\pi}^*_{\delta}$ with $\theta \in (1/2, 1]$ is consistent under (\ref{cond:theta_1}), no matter how strong the MAC dependence is. 

Although the above analyses are for the discretized version of $\hat \pi_\delta$, they provides important insight on the performance of the $\hat \pi_\delta$ family under dependence, which is also supported by extensive simulation studies in Section \ref{sec:simulation}.


\subsection{A new estimator for dependent variables} \label{sec:estimator}

Study in the previous section reveals very different effects of the MAC dependence on different estimators in the $\hat \pi_\delta$ family. 
When MAC dependence is relatively weak, estimators with $\theta\in [0, 1/2]$ may be more powerful as condition (\ref{cond:theta_05}) is less restrictive for smaller $\bar \rho_\Sigma$. On the other hand, when MAC dependence gets stronger, estimators with  $\theta\in (1/2, 1]$ could be more powerful as condition (\ref{cond:theta_1}) is free of $\bar \rho_\Sigma$. Motivated by these findings, we propose to construct a new estimator of the form 
\[
\hat \pi_{adap} = \max\{ \hat \pi_\delta, \delta \in \Delta\},
\]
where $\Delta$ is the set of $\delta(t)$ functions that render the most powerful estimators in different dependence scenarios. Note that such  $\delta(t)$ functions in  $\Delta$ all result in conservative estimators with the lower bound property as stated in Theorem \ref{thm:general_MR}. Therefore, the new estimator $\hat \pi_{adap}$ naturally inherit the lower bound property.  

The power of  $\hat \pi_{adap}$ depends on the candidate  $\delta(t)$ functions in $\Delta$. We consider $\delta(t) = [\bar \Phi(t)]^\theta$ with specific $\theta$ values. Informed by Theorem \ref{thm:discrete_0.5}, which is for all the estimators with $\theta \in [0, 1/2]$, we see that condition (\ref{cond:theta_05}) is less stringent with larger $\theta$. Therefore, we expect the estimator with $\theta = 1/2$ to be the most powerful candidate in this group. 
On the other hand,  based on the results in Theorem \ref{thm:discrete_1}, we expect the most powerful estimator in the group of estimators with $\theta \in (1/2, 1]$ to have $\theta=1$.  
The above analysis helps us narrow down to two candidates that are most powerful in their own $\theta$ groups.  
Comparing these two candidates, we see that  $\theta=1/2$ can result in a  more (or less) powerful estimator than  $\theta=1$ when MAC dependence is relatively weak (or strong) with $\bar \rho_{\Sigma} \le \pi/\sqrt{\log p}$ (or $\bar \rho_{\Sigma} > \pi/\sqrt{\log p}$). Since none of the candidates dominates the other under arbitrary covariance dependence, our new estimator is constructed as 
\begin{equation} \label{def:adap_pi}
\hat \pi_{adap} = \max\{\hat \pi_{0.5}, \hat \pi_{1}\},
\end{equation}
where $\hat \pi_{0.5}$ denotes $\hat \pi_{\delta}$ with $\delta = [\bar \Phi(t)]^{1/2}$, and $\hat \pi_{1}$ denotes $\hat \pi_{\delta}$ with $\delta = \bar \Phi(t)$. $\hat \pi_{adap}$ is likely to be comparable to 
$\hat \pi_{0.5}$ and outperform $\hat \pi_{1}$ under relatively weak dependence, and be comparable to $\hat \pi_{1}$ and outperform $\hat \pi_{0.5}$ when covariance dependence is strong.

\underline{Numerical Implementation.}
We conclude this section with additional notes on the numerical implementation of $\hat \pi_{adap}$. 
Specifically, we  simulate $(w_1, \ldots, w_p)$ following the joint null distribution of $Z_j$. When the joint null distribution is unknown in  real applications, $(w_1, \ldots, w_p)$ can often be simulated non-parametrically. For example, when $(Z_1, \ldots, Z_p)$ are a set of test statistics for associations between a set of explanatory variables and a response variable, a common practice to simulate $(w_1, \ldots, w_p)$ is by randomly shuffling only the sample of the response variable to remove the potential associations. More details for such permutation approaches can be found in \cite{westfall1993resampling}. Then, following (\ref{def:Vp}), we simulate $V_{p,0.5}$ and $V_{p,1}$ corresponding to $\theta=1/2$ and $1$ by taking the maximums over $t=w_1, \ldots, w_p$.  
Next, we set $\alpha = 0.1$ and generate $c_{p, 0.5}$ and $c_{p, 1}$ as the $(1-\alpha)$th quantile of 1000 replicates of $V_{p,0.5}$ and $V_{p,1}$, respectively. 
The simulated $c_{p, 0.5}$ and $c_{p, 1}$ are implemented to calculate  $\hat \pi_{0.5}$ and $\hat \pi_{1}$ as in (\ref{def:MR}) by taking the maximums over $t = z_1, \ldots z_p$, where  $z_1, \ldots z_p$ are the observed variables after inverse normal transformation. 
Finally, the new estimator $\hat \pi_{adap}$ is calculated by (\ref{def:adap_pi}). 

\section{Simulation Study} \label{sec:simulation}

In the following simulation examples, we consider six dependence structures: (a)-(d) are commonly observed correlation structures in literature and (e)-(f) are real correlation structures from genetic association studies. In all the examples, $\Sigma_{ii} = 1, i=1, \ldots, p$.

\begin{enumerate} [(a)]
	\item  {\bf Autoregressive}. $\Sigma_{ij} = r^{|i-j|}$ and $r = 0.9$. 
	\item {\bf Equal correlation}.  $\Sigma_{ij} = 0.5$ for $i\ne j$.
	\item {\bf Block correlation}. $\Sigma$ has square diagonal blocks. The off-diagonal elements in the blocks are 0.5, and the elements outside the blocks are zero.   
	\item {\bf Sparse correlation}. $\Sigma$ has nonzero elements randomly located. The data generation process is similar to Model 3 in \cite{cai2013two}. Let $\Sigma^* = (\sigma_{ij})$, where $\sigma_{ii} = 1$, $\sigma_{ij} = 0.9 *$ Bernoulli$(1, 0.1)$ for $i<j$ and $\sigma_{ji} = \sigma_{ij}$. Then $\Sigma = I^{1/2}(\Sigma^* + \delta I)/ (1+\delta) I^{1/2}$, where $\delta = |\lambda_{min} (\Sigma^*)| + 0.05$. 
	\item {\bf SNP correlation}. $\Sigma$ is the sample correlation matrix of the real SNP data on Chromosome 21 from 90 individuals in the
	International HapMap project. 
	\item {\bf Gene correlation}. $\Sigma$ is the sample correlation matrix of the real gene expression data from 71 individuals in a riboflavin production study.
\end{enumerate}

We generate test statistics $Z_1, \ldots, Z_p \sim N((\mu_1, \ldots, \mu_p), \Sigma)$ and set $p=2000$ for cases (a)-(d) above. Case (e) has $p=8657$, which is the number of SNPs in the dataset, and case (f) has  $p=4088$, which is the number of genes in the dataset.  Additional details of the datasets can be found in Section \ref{sec:application}. The block size in case (c) is set as $400\times 400$. 
$(\mu_1, \ldots, \mu_p)$ is a sparse vector with randomly located non-zero elements. We consider both relatively sparse signals with $\pi = 0.02$ and more dense signals with $\pi=0.1$. 

\subsection{MAC dependence effect on bounding sequences}

\begin{figure} [!h]
	\centering
	\caption{ Trends of MAC levels and bounding sequences.} \label{Fig:MACvsC}
	\vspace{-0.1in}
	\includegraphics[width=5in,height=3.5in]{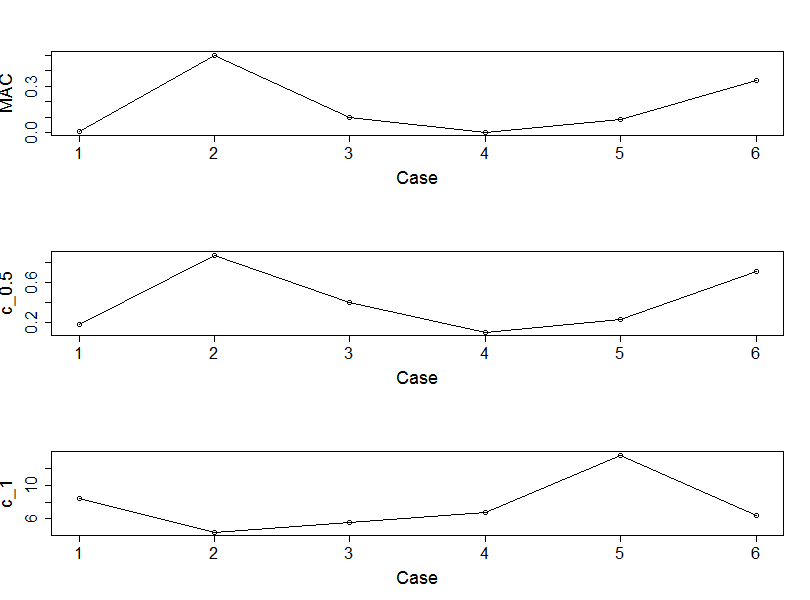}
\end{figure}

We first calculate the MAC levels as defined in (\ref{def:bar_rho}) for the dependence structures in (a)-(f) above and report the realized values of $c_{p, 0.5}$ and $c_{p, 1}$, which are generated by the procedure described at the end of Section \ref{sec:estimator}. Recall that  a larger value of $\bar\rho_\Sigma$ indicates stronger overall covariance dependence. 
It can be seen in Table \ref{tab:rho_c} that  $\bar\rho_\Sigma$ is fairly small for cases (a) and (d), moderately small for cases (c) and (e), and fairly large for cases (b) and (f). Moreover,
$c_{p, 0.5}$ seems to vary positively with $\bar\rho_\Sigma$, whereas $c_{p, 1}$ does not show such tendency. The numerical results are demonstrated more clearly in Figure \ref{Fig:MACvsC}, which seem to agree with the theoretical results in Theorem \ref{thm:discrete_0.5} and \ref{thm:discrete_1} about the MAC dependence effect on $c_{p, 0.5}$ and $c_{p, 1}$.

\begin{table}[!h] 
	\centering
	\caption{MAC levels and realized values of bounding sequences.} \label{tab:rho_c}
	\begin{tabular}{l c c c c c c}
		\hline
		\hline
		& Autocorr & Equal corr & Block corr &  Sparse corr  & SNP corr & Gene corr\\
		\hline
		$\bar \rho_{\Sigma}$ & 0.0095 & 0.5003 & 0.1003 & 0.0042 & 0.0869 & 0.3353 \\		
		$c_{p, 0.5}$ & 0.178 & 0.87 & 0.397 & 0.099 & 0.222 & 0.706 \\
		$c_{p,1}$ & 8.46 & 4.39 & 5.58 & 6.79 & 13.6 & 6.42\\
		\hline
	\end{tabular}
\end{table}

\newpage

\subsection{Comparison with existing methods} \label{sec:sim_compare}

We compare the new estimator $\hat \pi_{adap}$ with $\hat \pi_{0.5}$ and $\hat \pi_{1}$ from the estimator family as well as two other popular methods. 
Note that $\hat \pi_{0.5}$ is closely related to the optimal estimator in the original family of \cite{MR06} under independence, and $\hat \pi_{1}$ has been discovered to have good power for relatively sparse signals under independence. The two other existing methods are   $\hat \pi_{GW}$ developed in \cite{GW04} and $\hat \pi_{JC}$  developed in \cite{JC07}. These two existing methods have been studied for relatively dense signals under independence.

Besides various dependence structures in (a)-(f), we consider sparse and relatively dense signals with $\pi=0.02$ and $0.1$, respectively, and varying signal intensity with non-zero $\mu_j = 3, 4, 5, 6$. The comparisons are organized into two sets of examples. 

\begin{figure} [!h]
	\centering
	\caption{Comparison under autocorrelation. ``est\_0.05", ``est\_1", and ``est\_adapt" represent $\hat{\pi}_{0.5}$, $\hat \pi_1$, and $\hat \pi_{adap}$, respectively. The top row has $\pi=0.02$, and the bottom row has $\pi=0.1$. The true $\pi$ values are highlighted by the red horizontal lines.} \label{Fig:Auto}
	\vspace{-0.1in}
	\includegraphics[width=6.7in,height=1.78in]{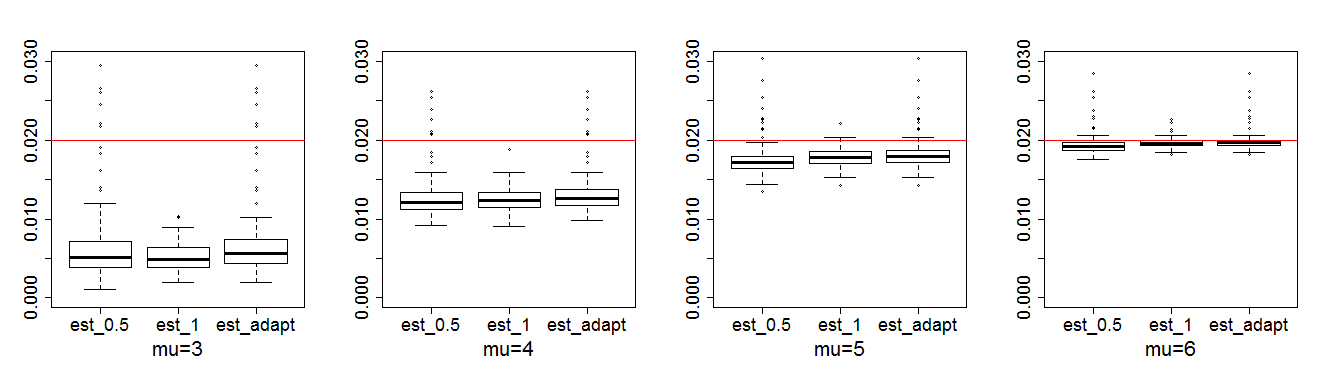}
	\includegraphics[width=6.7in,height=1.78in]{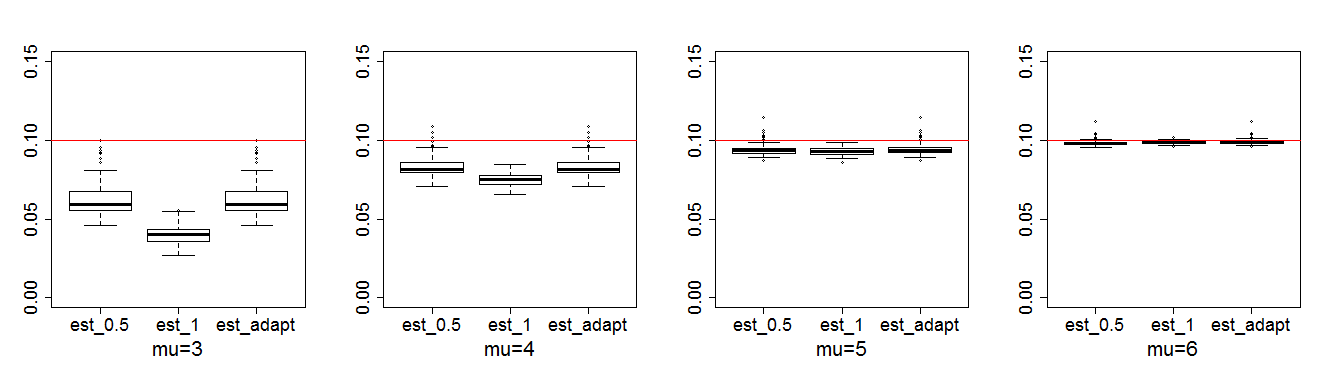}
\end{figure}

\begin{figure} [!h]
	\centering
	\caption{Comparison under equal correlation. Notations and symbols are the same as in Figure \ref{Fig:Auto}. } \label{Fig:Equal}
	\vspace{-0.1in}
	\includegraphics[width=6.5in,height=1.78in]{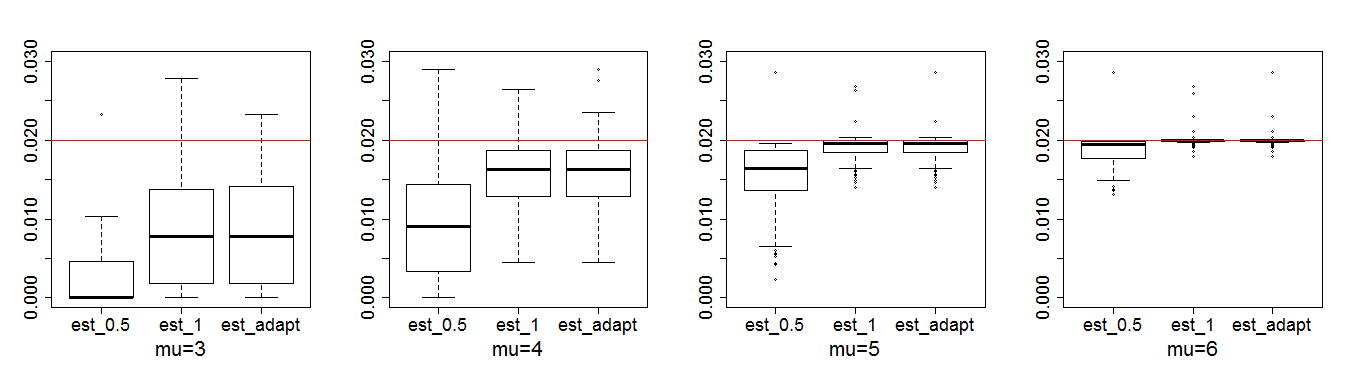}
	\includegraphics[width=6.5in,height=1.78in]{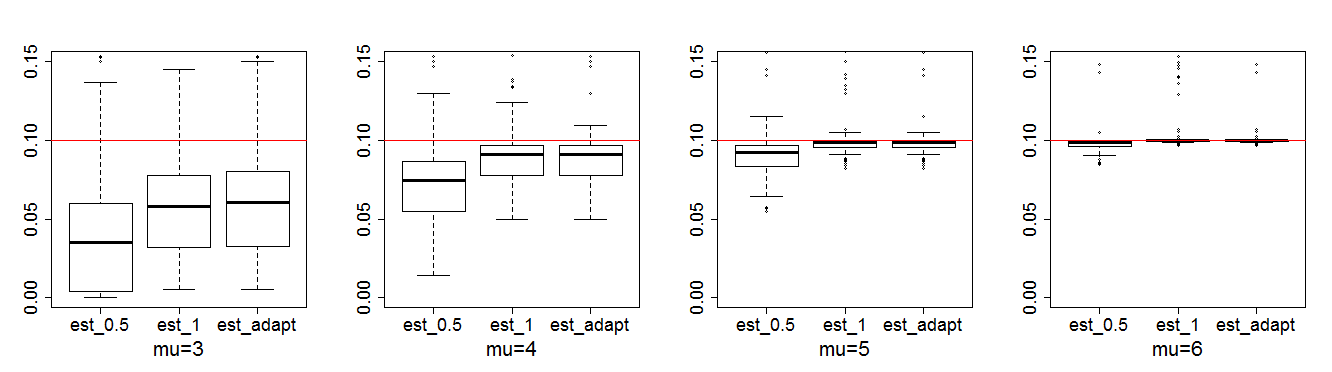}
\end{figure}

\begin{figure} [!h]
	\centering
	\caption{Comparison under block dependence.  Notations and symbols are the same as in Figure \ref{Fig:Auto}.} \label{Fig:Block400}
	\vspace{-0.1in}
	\includegraphics[width=6.5in,height=1.78in]{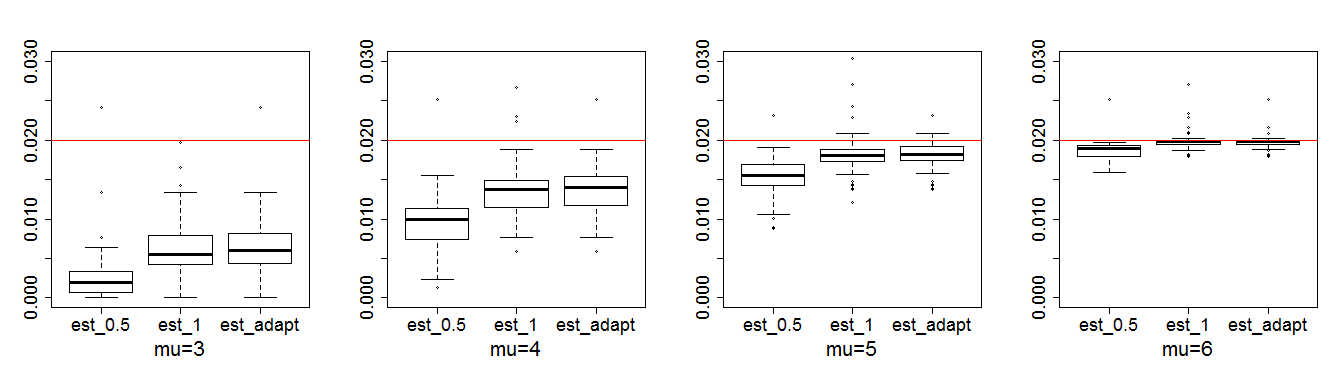}
	\includegraphics[width=6.5in,height=1.78in]{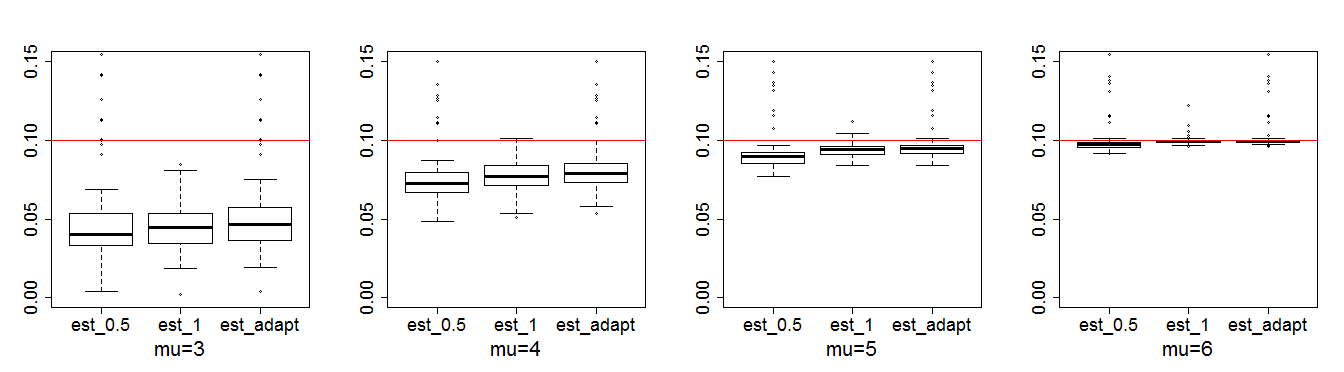}
\end{figure}

\begin{figure} [!h]
	\centering
	\caption{Comparison under sparse dependence. Notations and symbols are the same as in Figure \ref{Fig:Auto}.} \label{Fig:Sparse}
	\vspace{-0.1in}
	\includegraphics[width=6.5in,height=1.78in]{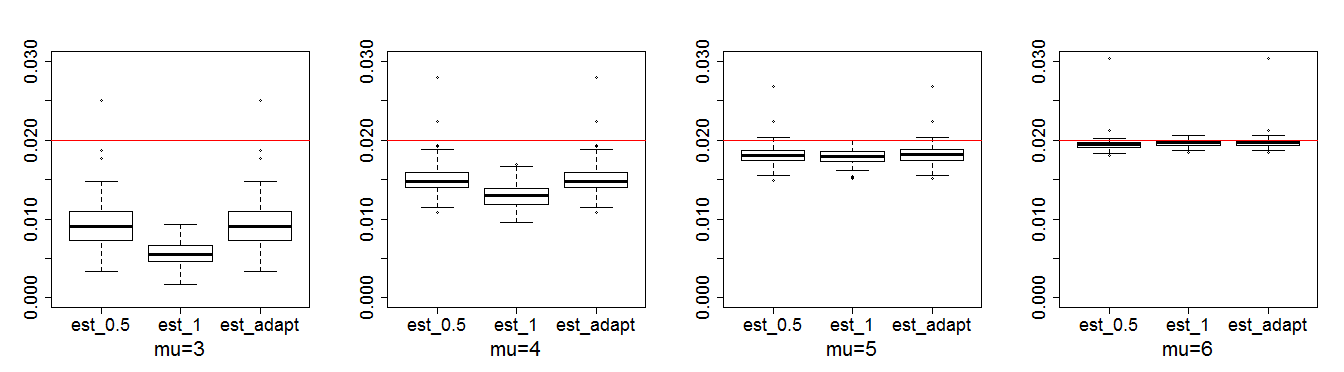}
	\includegraphics[width=6.5in,height=1.78in]{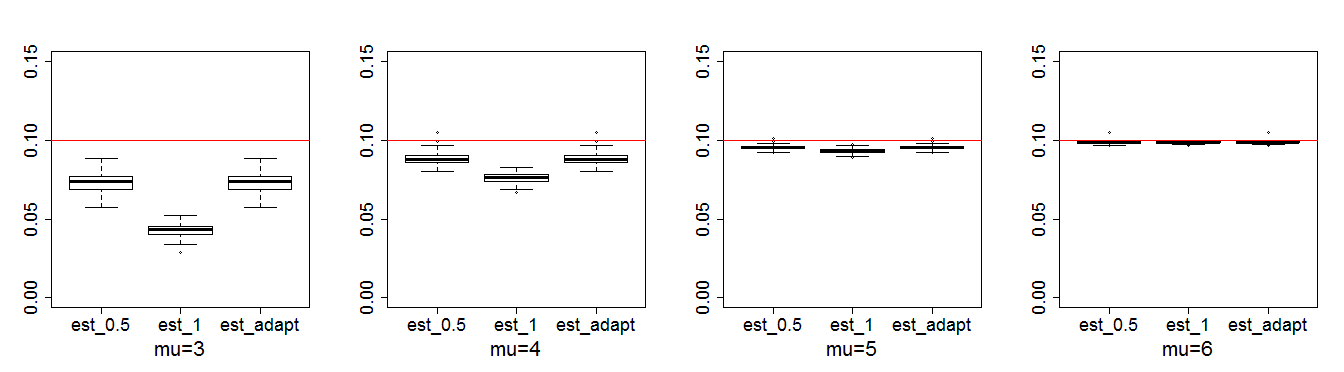}
\end{figure}

\begin{figure} [!h]
	\centering
	\caption{Comparison under SNP dependence.  Notations and symbols are the same as in Figure \ref{Fig:Auto}.} \label{Fig:SNP}
	\vspace{-0.1in}
	\includegraphics[width=6.5in,height=1.78in]{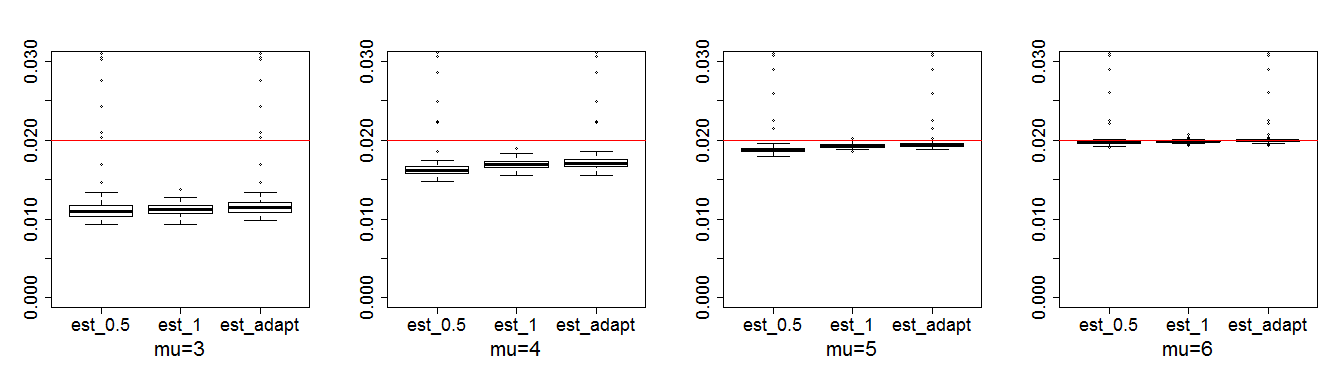}
	\includegraphics[width=6.5in,height=1.78in]{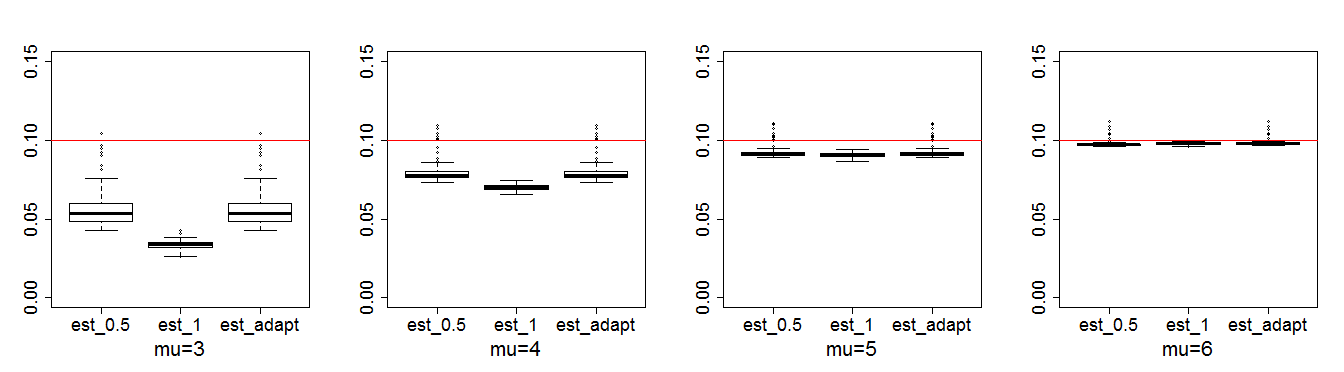}
\end{figure}

\begin{figure} [!h]
	\centering
	\caption{Comparison under gene dependence. Notations and symbols are the same as in Figure \ref{Fig:Auto}.} \label{Fig:Gene}
	\vspace{-0.1in}
	\includegraphics[width=6.5in,height=1.78in]{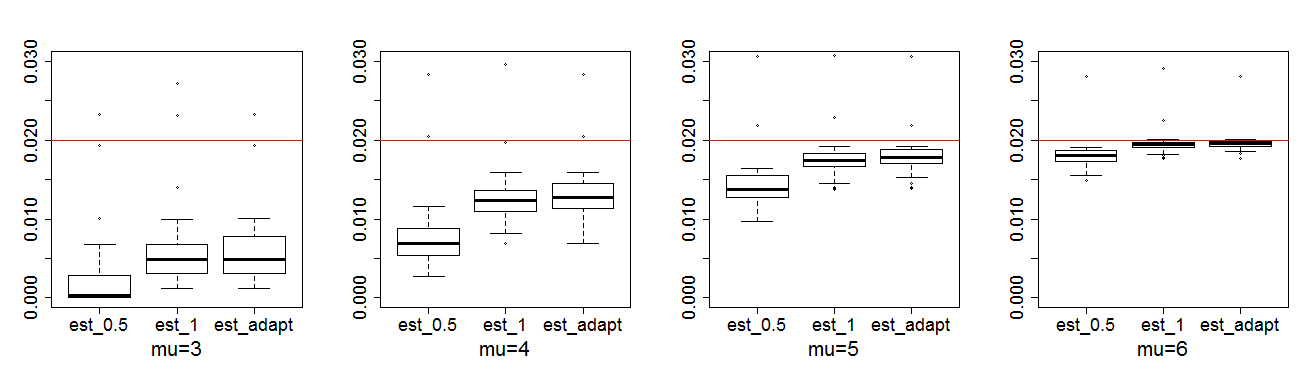}
	\includegraphics[width=6.5in,height=1.78in]{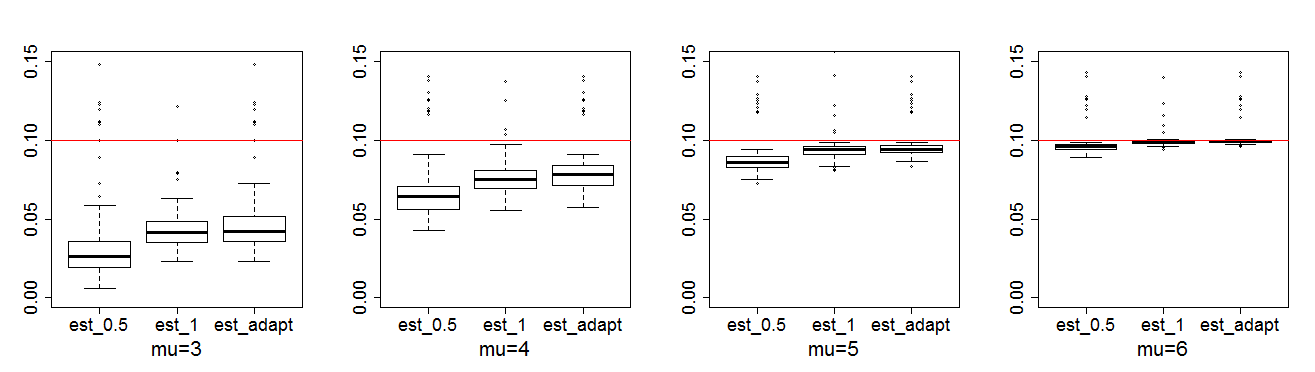}
\end{figure}

The first set of examples demonstrate the performances of $\hat \pi_{adap}$, $\hat \pi_{0.5}$, and $\hat \pi_{1}$ as they all possess the lower bound property.  
Recall the theoretical results in Section \ref{sec:estimation} that $\hat \pi_{0.5}$ may outperform $\hat \pi_{1}$ when dependence is weak enough or signals are less sparse, and $\hat \pi_{1}$ may perform better in the other scenarios. 
We observe such tendencies in Figure \ref{Fig:Auto}-\ref{Fig:Gene}. Specifically, the autocorrelation case (Figure \ref{Fig:Auto}) has small $\bar \rho_\Sigma=0.0095$. It shows that $\hat \pi_{0.5}$ has comparable results as those of $\hat \pi_{1}$ for small $\pi=0.02$, and outperforms $\hat \pi_{1}$ for larger $\pi=0.1$.  
The equal correlation case (Figure \ref{Fig:Equal}) has the largest $\bar \rho_\Sigma=0.5$. It shows that  $\hat \pi_{1}$ outperforms $\hat \pi_{0.5}$ for both $\pi=0.02$ and $0.1$. The block diagonal case (Figure \ref{Fig:Block400}) has 
moderate $\bar \rho_\Sigma=0.1$. It shows that $\hat \pi_{1}$ outperforms $\hat \pi_{0.5}$ for small $\pi$, and is comparable to  $\hat \pi_{0.5}$ for larger $\pi$. The sparse correlation case (Figure \ref{Fig:Sparse}) has the smallest $\bar \rho_\Sigma=0.0042$, we see that $\hat \pi_{0.5}$ outperforms $\hat \pi_{1}$ for both $\pi=0.02$ and $0.1$. The SNP correlation case (Figure \ref{Fig:SNP}) has $\bar \rho_\Sigma=0.0869$, which is moderately small. We see that $\hat \pi_{1}$ is slightly better for small $\pi$, and $\hat \pi_{0.5}$ is better for larger $\pi$. The gene correlation case (Figure \ref{Fig:Gene}) has $\bar \rho_\Sigma=0.3353$, which is fairly large. It shows that $\hat \pi_{1}$ outperforms $\hat \pi_{0.5}$ for both small and larger $\pi$. In all these examples, the new estimator $\hat \pi_{adap}$ always matches the winner of $\hat \pi_{0.5}$ and $\hat \pi_{1}$ and exhibits better adaptivity to different dependence structures and signal sparsity levels.

\pagebreak

\pagebreak

\pagebreak

\newpage

The second set of examples compares the performance of  $\hat \pi_{adap}$ with those of  $\hat \pi_{GW}$ from \cite{GW04} and $\hat \pi_{JC}$ from \cite{JC07}.
Table \ref{tab:compare_sparse} and \ref{tab:compare_dense} show that for the cases (a)-(f), $\hat \pi_{GW}$ and $\hat \pi_{JC}$ tend to over-estimate the true $\pi$ when dependence is strong (such as in (b) Equal correlation and (f) Gene correlation) or signals are sparse (such as for $\pi=0.02$). On the other hand, the performance of $\hat \pi_{adap}$ seems to be generally more accurate and stable over different dependence structures and $\pi$ values.

\begin{table}[!h] 
	\centering
	\caption{Mean values and standard deviations (in brackets) of $\hat \pi_{adap}$, $\hat \pi_{GW}$, and $\hat \pi_{JC}$  when signals are relatively spare with $\pi = 0.02$.} \label{tab:compare_sparse}
	\begin{tabular}{l l c c c c}
		\hline
		\hline
		Dependence & Method & $\mu=3$ & $\mu=4$ & $\mu=5$ & $\mu=6$ \\
		\hline
		Autocorr & $\hat \pi_{adap}$ & 0.020(0.003) & 0.020(0.003) & 0.020(0.002) & 0.021(0.002) \\		
		& $\hat \pi_{GW}$  & 0.014(0.024) & 0.015(0.024) & 0.014(0.024) & 0.014(0.024) \\
		& $\hat \pi_{JC}$  & 0.050(0.045) & 0.050(0.045) & 0.050(0.045) & 0.051(0.044) \\
		\hline
		Equal corr & $\hat \pi_{adap}$ & 0.034(0.051) & 0.036(0.050) & 0.038(0.049) & 0.038(0.049) \\		
		& $\hat \pi_{GW}$  & 0.194(0.274) & 0.194(0.273) & 0.194(0.273) & 0.194(0.274) \\
		& $\hat \pi_{JC}$  & 0.295(0.402) & 0.295(0.402) & 0.295(0.401) & 0.29690.402 \\
		\hline
		Block corr & $\hat \pi_{adap}$ & 0.012(0.019) & 0.018(0.017) & 0.023(0.016) & 0.024(0.016) \\		
		& $\hat \pi_{GW}$  & 0.060(0.091) & 0.060(0.091) & 0.059(0.091) & 0.060(0.091)  \\
		& $\hat \pi_{JC}$  & 0.115(0.143) & 0.115(0.142) & 0.116(0.141) & 0.118(0.140) \\
		\hline
		Sparse corr & $\hat \pi_{adap}$ & 0.009(0.003) & 0.015(0.022) & 0.018(0.001) & 0.020(0.001) \\		
		& $\hat \pi_{GW}$ & 0.003(0.006)  & 0.003(0.007) & 0.003(0.006) & 0.003(0.007) \\
		& $\hat \pi_{JC}$  & 0.030(0.021) & 0.031(0.020) & 0.032(0.020) & 0.033(0.020) \\
		\hline
		SNP corr & $\hat \pi_{adap}$ & 0.034 (0.051) & 0.036(0.050) & 0.038(0.049) & 0.038(0.049) \\		
		& $\hat \pi_{GW}$  & 0.194(0.274) & 0.194(0.273) & 0.194(0.273) & 0.194(0.274) \\
		& $\hat \pi_{JC}$  & 0.295(0.402) & 0.295(0.402) & 0.295(0.401) & 0.296(0.402) \\
		\hline
		Gene corr & $\hat \pi_{adap}$ & 0.019(0.038) & 0.026(0.037)& 0.030(0.035)& 0.032(0.035) \\		
		& $\hat \pi_{GW}$  & 0.151(0.187) & 0.151(0.188) & 0.151(0.188) & 0.151(0.187) \\
		& $\hat \pi_{JC}$  & 0.257(0.308) & 0.256(0.307)& 0.257(0.307) & 0.258(0.306)\\
		\hline
	\end{tabular}
\end{table}

\begin{table}[!h] 
	\centering
	\caption{Mean values and standard deviations (in brackets) of $\hat \pi_{adap}$, $\hat \pi_{GW}$, and $\hat \pi_{JC}$ when signals are relatively dense with $\pi = 0.1$.} \label{tab:compare_dense}
	\begin{tabular}{l l c c c c}
		\hline
		\hline
		Dependence & Method & $\mu=3$ & $\mu=4$ & $\mu=5$ & $\mu=6$ \\
		\hline
		Autocorr & $\hat \pi_{adap}$ &  0.063(0.011)  & 0.083(0.007) & 0.094(0.004) & 0.100(0.002)  \\		
		& $\hat \pi_{GW}$  & 0.078(0.029) & 0.086(0.026) & 0.088(0.023) & 0.088(0.024) \\
		& $\hat \pi_{JC}$  &  0.119(0.046) & 0.124(0.043) & 0.127(0.043) & 0.125(0.041) \\
		\hline
		Equal corr & $\hat \pi_{adap}$ &  0.067(0.053) & 0.098(0.042) & 0.110(0.038) & 0.113(0.036) \\		
		& $\hat \pi_{GW}$  &  0.233(0.263) & 0.246(0.256) & 0.250(0.254) & 0.250(0.253) \\
		& $\hat \pi_{JC}$  &  0.333(0.394) & 0.336(0.377) & 0.344(0.374) & 0.349(0.373) \\
		\hline
		Block diag & $\hat \pi_{adap}$ &  0.052(0.026) & 0.082(0.018) &  0.097(0.013) & 0.102(0.012)  \\		
		& $\hat \pi_{GW}$  &   0.114(0.094) & 0.126(0.089) & 0.128(0.085) & 0.129(0.086) \\
		& $\hat \pi_{JC}$  &  0.173(0.144) & 0.179(0.135) & 0.183(0.131) & 0.184(0.131) \\
		\hline
		Sparse corr & $\hat \pi_{adap}$ & 0.073(0.006) & 0.088(0.004) & 0.096(0.001)  & 0.099(0.001) \\		
		& $\hat \pi_{GW}$ &  0.069(0.012) & 0.075(0.010) & 0.077(0.009) & 0.077(0.009)  \\
		& $\hat \pi_{JC}$  &  0.106(0.019) & 0.108(0.019) & 0.110(0.019) & 0.110(0.018) \\
		\hline
		SNP corr & $\hat \pi_{adap}$ & 0.057(0.013) & 0.080(0.007) & 0.093(0.004) &  0.098(0.002) \\		
		& $\hat \pi_{GW}$  &   0.100(0.043) & 0.107(0.038) & 0.110(0.036) & 0.111(0.035) \\
		& $\hat \pi_{JC}$  &   0.129(0.070) & 0.133(0.066) & 0.137(0.064) & 0.138(0.063) \\
		\hline
		Gene corr & $\hat \pi_{adap}$ &  0.055(0.038) & 0.087(0.033) & 0.102(0.028) & 0.107(0.026) \\		
		& $\hat \pi_{GW}$  &  0.202(0.181)  & 0.213(0.177) & 0.216(0.173) & 0.217(0.172) \\
		& $\hat \pi_{JC}$  &  0.301(0.302) & 0.304(0.292)  & 0.312(0.287) & 0.314(0.284) \\
		\hline
	\end{tabular}
\end{table}
		
\newpage
		
\section{Real Application} \label{sec:application}

We apply the proposed method to two real datasets. The first dataset is from an eQTL study with the goal 
to identify SNPs that  potentially govern the expression of gene CCT8 on chromosome 21. This gene has been found to be relevant to Down Syndrome \citep{bradic2011penalized, fan2012estimating}. We obtain the SNP data of unaffected subjects from the International HapMap project (\url{http://zzz.bwh.harvard.edu/plink/res.shtml#hapmap}) and the gene expression data from \url{ ftp://ftp.sanger.ac.uk/pub/genevar/}. 
Our data includes 90 samples from Asian population ($45$ Japanese in Tokyo, Japan (JPT), and $45$ Han Chinese in Beijing). 
We consider SNPs without missing values, which results in 8657 candidate SNPs. 

\begin{figure} [!h]
	\centering
	\caption{Histogram of test statistics for eQTL analysis} \label{Fig:hist_SNP}
	\vspace{-0.1in}
	\includegraphics[width=4.5in,height=2.5in]{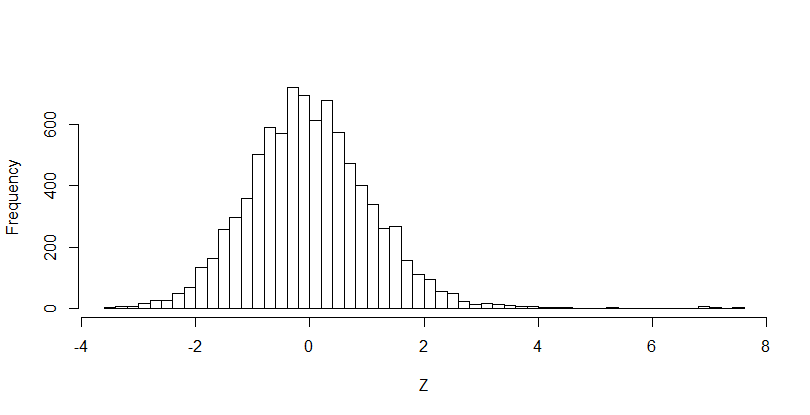}
\end{figure}

\begin{figure} [!h]
	\centering
	\caption{Heatmap of the absolute value of correlations for 50 SNPs in Hapmap data.} \label{Fig:heatmap_SNP}
	\vspace{-0.1in}
	\includegraphics[width=6in,height=3in]{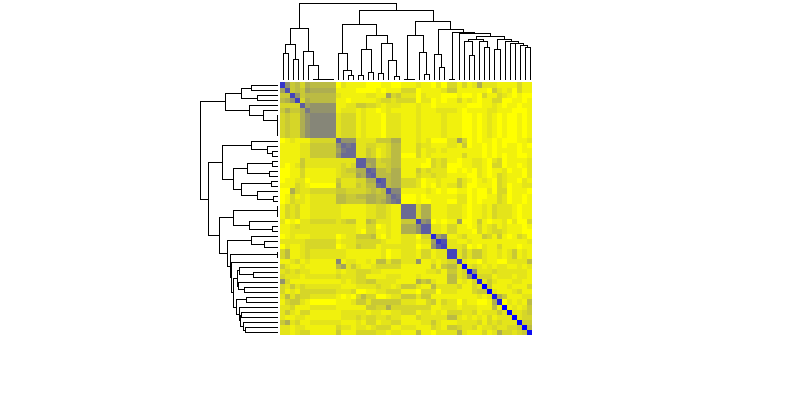}
\end{figure}

Test statistics for the associations between each SNP and the expression level of CCT8 are derived by marginal linear regression as in \cite{bradic2011penalized} and \cite{fan2012estimating}. Histogram of the test statistics is presented in \autoref{Fig:hist_SNP}, where the long and thin right tail indicates possibly a small proportion of signals with positive signal effects.  The correlation matrix of the test statistics, which is the same as the correlation matrix of the SNPs, has the MAC level of $\bar \rho_\Sigma = 0.087$. The heatmap of the correlation matrix of the first 50 SNPs is illustrated in \autoref{Fig:heatmap_SNP}.  

We apply the proposed estimator $\hat \pi_{adap}$ and the existing estimators  $\hat \pi_{GW}$ and $\hat \pi_{JC}$ to the dataset. The results are $\hat \pi_{adap} = 0.0016$, $\hat \pi_{GW} = 0.0068$, and $\hat \pi_{JC} = 0.0186$. Consequently, the estimated numbers of relevant SNPs are  $14$, $60$, and $161$ by the three methods, respectively. 
The dependence structure of the real data has been adopted in simulation study to help with justifying the real data results. 
The MAC level and the realized values of the bounding sequences $c_{p, 0.5}$ and $c_{p,1}$ have been demonstrated in Table \ref{tab:rho_c}, case (e).  
Simulation results that are most relevant are presented in the top row of \autoref{Fig:SNP}  and the sixth row of \autoref{tab:compare_sparse}. The real application results here seem to be consistent with the findings in simulation, 
where $\hat \pi_{adap}$ is much smaller than $\hat \pi_{GW}$ and $\hat \pi_{JC}$ and is closer to the true $\pi$.

\begin{figure} [!h]
	\centering
	\caption{Histogram of test statistics for gene expression association} \label{Fig:hist_gene}
	\vspace{-0.1in}
	\includegraphics[width=4.5in,height=2.5in]{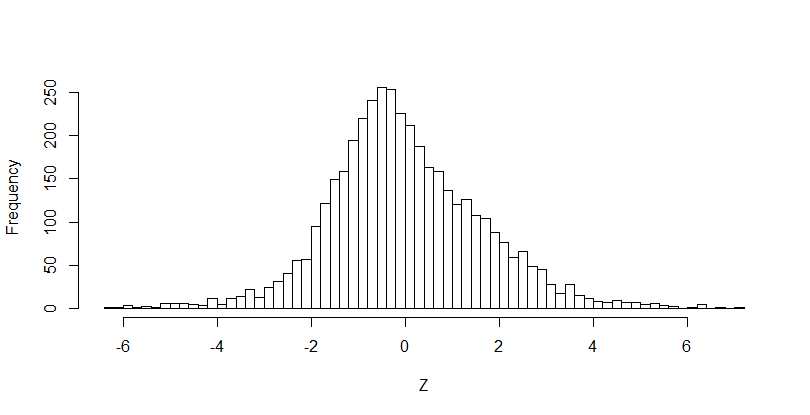}
\end{figure}

\begin{figure} [!h]
	\centering
	\caption{Heatmap of the absolute value of correlations for 50 gene expressions in riboflavin production study.} \label{Fig:Heatmap_gene}
	\vspace{-0.1in}
	\includegraphics[width=6in,height=3in]{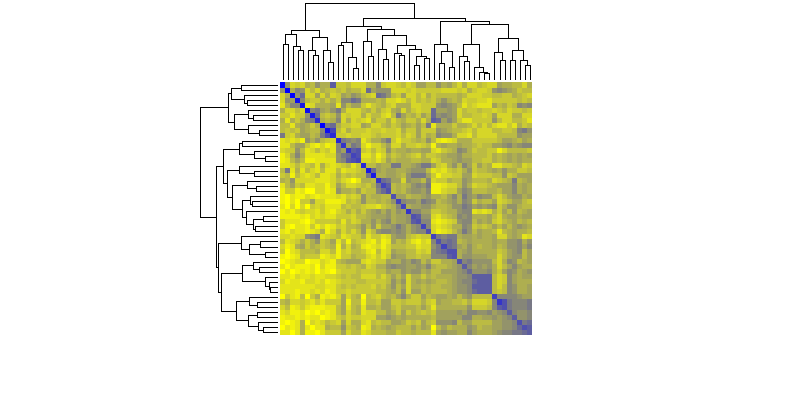}
\end{figure}

The second real application example has microarray data from a study on riboflavin (vitamin B$_2$) production in bacillus subtilis. This dataset is available at \url{https://www.annualreviews.org/doi/suppl/10.1146/annurev-statistics-022513-115545} and has been studied in  \cite{Buhlmann2014}. The dataset includes the expression levels of 4088 genes and the logarithm of riboflavin production rate of $71$ individuals. Marginal regression coefficients are used as test statistics for associations between genes and riboflavin production. The histogram of the test statistics is presented in \autoref{Fig:hist_gene}, which suggests a larger signal proportion than that in the first real data example. 
\autoref{Fig:Heatmap_gene}  shows the heatmap of the correlation matrix of the first 50 genes, which indicates a more complicated dependence structure.  The MAC level of the genes is $\bar \rho_{\Sigma} = 0.335$, which is fairly large. The realized value of the bounding sequences $c_{p, 0.5}$ and $c_{p,1}$ have been demonstrated in Table \ref{tab:rho_c}, case (f).


In this example, we have $\hat \pi_{adap} = 0.064$, $\hat \pi_{GW} = 0.258$, and $\hat \pi_{JC} = 0.374$, which corresponds to $261$, $1053$, and $1530$ relevant genes. This real dependence structure has been adopted in simulation study. The most relevant simulation results are presented in the bottom row of Figure \ref{Fig:Gene} and the last row of Table \ref{tab:compare_dense}, which help us justify the real application results.  We can see that the results here seem to be consistent with the findings in simulation, where $\hat \pi_{adap}$ is much smaller than $\hat \pi_{GW}$ and $\hat \pi_{JC}$, and is closer to the true $\pi$.

\section{Conclusion and Discussion} \label{sec:conclusion}

Estimating the proportion of sparse signals is notoriously difficult when dealing with large-scale data with complex dependence structures. 
In this paper, we define the MAC level to measure arbitrary covariance dependence and explicate the joint effect of MAC dependence and signal sparsity on a family of estimators.
We find that no single estimator in the family is most powerful under different MAC dependence levels. We identify candidate estimators that are most powerful in different dependence scenarios and develop a new estimator that better adapts to arbitrary covariance dependence. The new estimator inherits the lower bound property of the family and provides a conservative estimate under very general conditions. This property is valuable in real applications as it requires no conditions on the unknown signals. 
Moreover, the new estimator is more powerful than representative members in the estimator family and compares favorably to other popular methods in extensive numerical examples including weak to strong covariance dependence and real dependence structures from genetic associations studies.
By pushing the frontier of high-dimensional sparse inference to better accommodate the complex data structures in real applications, we expect the impact of the proposed research to be far-reaching. 



\section*{Appendix} 

This section presents the proofs of Theorem \ref{thm:general_MR}, Theorem \ref{thm:discrete_0.5}, and Theorem \ref{thm:discrete_1}. The symbol $C$ denotes a genetic, finite constant whose values can be different at different occurrences.

\subsection{Proof of Theorem \ref{thm:general_MR}}

We first show $P(\hat \pi_\delta < \pi) \ge 1-\alpha$. Let $Z_j^0 = Z_j$ for $j \in I_0$ and $Z_j^1 = Z_j$ for $j \in I_1$. Denote $p_0=|I_0|$  and $s = |I_1|$. Then 
\begin{eqnarray*}
	\bar F_{p}(t) & = & p^{-1}\sum_{j\in I_{0}}1_{\{|Z_{j}^{0}|>t\}} + p^{-1} \sum_{j\in I_{1}}1_{\{|Z_{j}^{1}|>t\}} \le p^{-1}\sum_{j\in I_{0}}1_{\{|Z_{j}^{0}|>t\}} + p^{-1} s \\
	& = & (1-\pi) p_0^{-1} \sum_{j\in I_{0}}1_{\{|Z_{j}^{0}|>t\}} + \pi. 
\end{eqnarray*}
Consequently, 
\begin{eqnarray*}
	P(\hat{\pi}_\delta>\pi ) &\leq &P\left( \sup_{t>0} \left\{
	(1-\pi )\left( p_{0}^{-1}\sum\nolimits_{j\in I_{0}}1_{\{|Z_{j}^{0}|>t\}}-2\bar{\Phi}(t)\right) -c_{p,\delta}\delta(t)\right\} >0\right)
	\\
	&\leq &P\left( \sup_{t>0}\left\{ p_{0}^{-1}\sum\nolimits_{j\in
		I_{0}}1_{\{|Z_{j}^{0}|>t\}}-2\bar{\Phi}(t)-c_{p_0, \delta}\delta(t)\right\} >0\right) \\
	&\leq &P\left( V_{p_0}>c_{p_{0}, \delta}\right) = \alpha,
\end{eqnarray*}%
where the second and last inequalities are by properties (a) and (b) of $c_{p,\delta}$, respectively. The claim $P(\hat \pi_\delta < \pi) \ge 1-\alpha$ follows. 

Next, we show $P(\hat \pi_\delta > (1-\epsilon) \pi) \to 1$. Because  $\hat{\pi}_\delta >\bar{F}_{p}(t)-2\bar{\Phi}(t)-c_{p,\delta}
\delta\left( t\right)$ for any $t>0$ and 
\begin{equation*}
\bar{F}_{p}(t)={\frac{1-\pi }{p_{0}}}\sum_{j\in I_{0}}1_{\{|Z_{j}^{0}|>t\}} + {\frac{\pi }{s}}\sum_{j\in I_{1}}1_{\{|Z_{j}^{1}|>t\}},
\end{equation*}%
then
\begin{eqnarray} \label{0}
\frac{\hat{\pi}_\delta}{\pi }-1 & > &-\frac{1}{\pi }c_{p,\delta}\delta(t) + \frac{1-\pi }{\pi }\left( p_{0}^{-1}\sum\nolimits_{j\in I_{0}}1_{\{|Z_{j}^{0}|>t\}} -2\bar{\Phi}(t)\right) \nonumber \\   
& + & \left( {s}^{-1}\sum\nolimits_{j\in
	I_{1}}1_{\{|Z_j^1|>t\}}-1\right) -2\bar{\Phi}(t)
\end{eqnarray}%
for any $t >0$. Now, set $t$ in (\ref{0}) at $\tau$ such that $\tau\gg 1$ and $\delta(\tau) \ll \pi / c_{p,\delta}$. 
We will show that each term on the right hand side of (\ref{0}) at $t=\tau$ is of $o_p(1)$. 

First, by the condition $\delta(\tau) \ll \pi / c_{p,\delta}$, we have the first term $A_1 = -c_{p,\delta}\delta(\tau)/\pi = o(1)$. 

Consider the second term $A_2 = \pi^{-1} (1-\pi)\left( p_{0}^{-1}\sum\nolimits_{j\in I_{0}}1_{\{|Z_{j}^{0}|>\tau\}} -2\bar{\Phi}(\tau)\right)$ in (\ref{0}). The following lemma is proved in \autoref{sec:Lemma1proof}. 

\begin{lemma} \label{lemma:Z^0(tau)}
	For any  $\tau$ such that $\tau\gg 1$ and $\delta(\tau) \ll \pi / c_{p,\delta}$,  we have
	\[
	\pi^{-1} \left( p_{0}^{-1}\sum\nolimits_{j\in I_{0}}1_{\{|Z_{j}^{0}|>\tau\}} -2\bar{\Phi}(\tau)\right) = o_p(1).
	\]
\end{lemma}
\autoref{lemma:Z^0(tau)} and the condition $\pi= o(1)$ are enough to show $A_2 = o_p(1)$.

For the third term $A_3 = {s}^{-1}\sum\nolimits_{j\in
	I_{1}}1_{\{|Z_j^1|>\tau\}}-1$ in (\ref{0}), we have
\begin{eqnarray*}
	P(|A_3| > a) & = & P(1-{s}^{-1}\sum\nolimits_{j\in
		I_{1}}1_{\{|Z_j^1|>\tau\}} > a) \\
	& \le & a^{-1} (1- P(|Z_j^1| >\tau)) \\
	& = & a^{-1} \left(G(\tau) -  G(-\tau)\right) = o(1) 
\end{eqnarray*}
for any fixed $a >0$, where the third step is by $Z_j^1 \sim G$ for $j\in I_1$, and the last the step is by the condition $G(\tau) \to 0$ or $G(-\tau) \to 1$.

Last but not least, the forth term in (\ref{0}): $A_4 =-2\bar{\Phi}(\tau) = o(1)$ given $\tau \gg 1$. 

Summarizing the above gives the desired result $P(\hat \pi_\delta/\pi > 1-\epsilon) \to 1$.

\subsection{Proof of \autoref{lemma:Z^0(tau)}} \label{sec:Lemma1proof}

Recall the definitions of $V_{p, \delta}$ in (\ref{def:Vp}) and
\[
P\left(\sup_{t>0}{\frac{|p^{-1}\sum_{j=1}^{p}1_{\{|W_{j}|>t\}}-2\bar{\Phi}(t)|}{\delta\left(t\right) }} > c_{p,\delta} \right) < \alpha.
\]
where $W_1, \ldots, W_p \sim N_p(0, \Sigma)$. Then, for $t = \tau$,  
\[
P\left({\frac{|p^{-1}\sum_{j=1}^{p}1_{\{|W_{j}|>\tau\}}-2\bar{\Phi}(\tau)|}{\pi }} > \frac{c_{p,\delta} \delta(\tau)}{\pi} \right) <\alpha.
\]
Given  $c_{p,\delta} \delta(\tau) / \pi = o(1)$, we have
\begin{equation} \label{0.1}
{\frac{|p^{-1}\sum_{j=1}^{p}1_{\{|W_{j}|>\tau\}}-2\bar{\Phi}(\tau)|}{\pi }} = o_p(1).
\end{equation}
Decompose the left hand side above as 
\begin{eqnarray*}
	\pi^{-1} \left|p^{-1}\sum\nolimits_{j=1}^{p}1_{\{|W_{j}|>\tau\}}-2\bar{\Phi}(\tau)\right| & \ge & \pi^{-1}\left|p_0^{-1}\sum\nolimits_{j \in I_0}1_{\{|W_{j}|>\tau\}}-2\bar{\Phi}(\tau)\right| \\
	& - & \pi^{-1} \left|p^{-1}\sum_{j=1}^p 1_{\{|W_j|>\tau\}}- p_0^{-1}\sum\nolimits_{j \in I_0} 1_{\{|W_j|>\tau\}} \right|
\end{eqnarray*}
For the second term on the right hand side, 
\begin{eqnarray} \label{0.2}
\pi^{-1} \left|p^{-1}\sum\nolimits_{j=1}^p 1_{\{|W_j|>\tau\}}- p_0^{-1}\sum\nolimits_{j \in I_0} 1_{\{|W_j|>\tau\}} \right| & = & \pi^{-1}\left|p^{-1} \sum\nolimits_{j \in I_1} 1_{\{|W_j|>\tau\}} - \pi p_0^{-1} \sum\nolimits_{j \in I_0} 1_{\{|W_j|>\tau\}}\right| \nonumber\\
& \le & s^{-1} \sum\nolimits_{j \in I_1} 1_{\{|W_j|>\tau\}} + p_0^{-1} \sum\nolimits_{j \in I_0} 1_{\{|W_j|>\tau\}}.
\end{eqnarray}
Since $\tau \gg 1$, both $s^{-1} \sum\nolimits_{j \in I_1} 1_{\{|W_j|>\tau\}} = o_p(1)$ and $p_0^{-1} \sum\nolimits_{j \in I_0} 1_{\{|W_j|>\tau\}} = o_p(1)$ by Markov's inequality. Combining this with (\ref{0.1}) and (\ref{0.2}) gives 
\[
\pi^{-1}\left|p_0^{-1}\sum\nolimits_{j \in I_0}1_{\{|W_{j}|>\tau\}}-2\bar{\Phi}(\tau)\right| = o_p(1). 
\]
Now, because the joint distribution of $Z_j^0, j \in I_0$, is the same as the joint distribution of $W_j, j\in I_0$, claim in Lemma \ref{lemma:Z^0(tau)} follows. 

\subsection{Proof of Theorem \ref{thm:discrete_0.5}} \label{sec:proof_discrete0.5}

First, we show that given  $\delta(t) = [\bar \Phi(t)]^{\theta}, \theta \in [0,1/2]$, 
$c^*_{p, \delta} = C \sqrt{\bar \rho_\Sigma (\log p)^{\theta+1/2}}$, with a large enough constant $C$, satisfies properties (a) $p c^*_{p, \delta} > p_0 c^*_{p_0, \delta}$ and (b)  $P(V^*_{p,\delta} > c^*_{p, \delta}) < \alpha$ for all  $p$.  

Consider property (a).  Define $\Sigma_0$ as the covariance matrix of $W_j, j \in I_0$ and 
\begin{equation*} \label{def:bar_rho0}
\bar \rho_{\Sigma_0} =  \sum _{i\in I_0} \sum_{j \in I_0} |\Sigma_{ij}| / p^2_0. 
\end{equation*}
It can be shown that 
\[
\bar \rho_{\Sigma} > {1 \over p^2} \sum _{i\in I_0} \sum_{j \in I_0} |\Sigma_{ij}| = {(1-\pi)^2 \over p_0^2} \sum _{i\in I_0} \sum_{j \in I_0} |\Sigma_{ij}|  = (1-\pi)^2 \bar \rho_{\Sigma_0}.
\] 
Then it follows that 
\[
c^*_{p, \delta} > C(1-\pi) \sqrt{\bar \rho_{\Sigma_0} (\log p)^{\theta+1/2}} > (1-\pi) c^*_{p_0, \delta} = (p_0/p) c^*_{p_0, \delta},
\]
and property (a) is verified. 

Next consider property (b). 
By Chebyshev's inequality and direct calculation,
\begin{eqnarray*}
	P(V_{p, \delta}^*>c_{p, \delta}^*) &\leq &  \left( c_{p, \delta}^*\right) ^{-2} \mathsf{Var} (V_{p, \delta}^*) \le  \left( c_{p, \delta}^*\right) ^{-2} E ([V_{p, \delta}^*]^2) \\
	& = & \left( c_{p, \delta}^*\right) ^{-2} E \left[\max_{t\in \mathbb{T}} \left(\frac{|\bar W_p(t)-2\bar{\Phi}(t)|}{ [\bar \Phi(t)]^\theta } \right)^2 \right]
\end{eqnarray*}	
Let $A(t) =  [\bar \Phi(t)]^{-2\theta} (\bar W_p(t)-2\bar{\Phi}(t))^2$. It can be shown that 
\begin{eqnarray*}	
E \left(\max_{t\in \mathbb{T}} A(t) \right) & = & \int_0^\infty P(\max_{t\in \mathbb{T}} A(t) >c) d c \le \int_0^\infty \sum_{t \in \mathbb{T}} P(A(t) >c) dc \\
& = & \sum_{t \in \mathbb{T}} E [A(t)] \le C \sqrt{\log p} \cdot \max_{t\in \mathbb{T}} E [A(t)] \\
& = & C \sqrt{\log p}  \cdot \max_{t\in \mathbb{T}} \left\{[\bar \Phi(t)]^{-2\theta}  \mathsf{Var}\left(\bar W_p(t) \right) \right\}
\end{eqnarray*}
The following lemma provides the order of $\mathsf{Var} (\bar W_p(t))$. 

\begin{lemma} \label{lemma:Var(bar_W)} 
	For $W_1, \ldots, W_p \sim N_p(0, \Sigma)$ and $\bar \rho_{\Sigma}$  in (\ref{def:bar_rho}), 
	\begin{equation} \label{eq:var_barW}
	\mathsf{Var}\left(\bar W_p(t) \right) = O\left(\bar \rho_{\Sigma} \cdot e^{-t^2/2} \right).
	\end{equation}
\end{lemma}

Therefore, 
\begin{eqnarray*}
[\bar \Phi(t)]^{-2\theta} \cdot \mathsf{Var}\left(\bar W_p(t) \right) & \le & C [\bar \Phi(t)]^{-2\theta} \cdot \bar \rho_{\Sigma} \cdot e^{-t^2/2} \\
& \le &  C \left({t \over e^{-t^2/2}}\right)^{2\theta} \cdot \bar \rho_{\Sigma} \cdot e^{-t^2/2} \le C  (\log p)^{\theta} \cdot \bar \rho_{\Sigma} \cdot e^{(\theta-1/2)t^2} \le C \bar \rho_{\Sigma} \cdot (\log p)^{\theta}
\end{eqnarray*}
where the first step above is by Lemma  \ref{lemma:Var(bar_W)}, the second step is by Mill's ratio, the third step is by $t \in \mathbb{T}$, and the last step is by $\theta \in [0, 1/2]$. Combining the above, we have 
\[
P(V_{p, \delta}^*>c_{p, \delta}^*) \le C \left( c_{p, \delta}^*\right) ^{-2} \cdot \bar \rho_{\Sigma} \cdot (\log p)^{\theta+1/2} ,
\]
and $c^*_{p, \delta} = O(\sqrt{\bar \rho_\Sigma (\log p)^{\theta+1/2}})$ follows. 

Next, we demonstrate the upper bound property of $\hat \pi_\delta^*$.
Denote 
\[
B_p = \bar \Phi^{-1}\left( {\pi^{1/\theta} \over \bar \rho_{\Sigma}^{1/(2\theta)} (\log p)^{(\theta+1/2)/(2\theta)}} \right).
\]
Let $\tau = (A_p + B_p)/2$. Then, $A_p \gg 1$ and condition (\ref{cond:theta_05}) imply that $\tau \gg 1$ and $\tau - B_p \to \infty$, which further imply
\[
[\bar \Phi(\tau)]^\theta  \ll \pi / c^*_{p, \delta}.
\]
On the other hand,
\[
G_p(\tau) = \Phi(\tau-A_p) = \Phi(-(A_p-B_p)/2) = o(1),
\]
where the last step is by condition (\ref{cond:theta_05}). The rest is straightforward by applying Theorem \ref{thm:general_MR}. 

\subsection{Proof of Lemma \ref{lemma:Var(bar_W)}}
\[
\mathsf{Var}\left(\bar W_p(t) \right) = p^{-2}\sum_{j=1}^{p}Var(1_{\{|W_j| > t\}})+p^{-2}\sum_{i\neq j} Cov(1_{\{|W_i| >t\}},1_{\{|W_j| >t\}}).
\]
By Mill's ratio,
\begin{align*} \label{1.1}
p^{-2}\sum_{j=1}^{p}Var(1_{\{|W_j| > t\}})\leq p^{-1} 2\Bar{\Phi}(t)(1-2\Bar{\Phi}(t)) \le C p^{-1} e^{-t^2/2}. 
\end{align*}
For $p^{-2}\sum_{i\neq j} Cov(1_{\{|W_i| >t\}},1_{\{|W_j| >t\}})$, we have
\[
Cov(1_{\{|W_i| >t\}},1_{\{|W_j| >t\}})  =  4 \int_{-\infty}^{t}\int_{-\infty}^{t}f(x,y)dxdy - 4\int_{-\infty}^{t}\phi(x)dx\int_{-\infty}^{t}\phi(y)dy \le  C |\Sigma_{ij}| e^{-t^2/2},
\]
where the last step follows from Corollary 2.1 in \cite{li2002normal}. Combining the above with the definition of $\bar \rho_{\Sigma}$ results in (\ref{eq:var_barW}).

\subsection{Proof of Theorem \ref{thm:discrete_1}}

First, it is easy to see that $c^*_{p, \delta} = C \sqrt{\log p}$ satisfies property (a)  $p c_{p, \delta} > p_0  c_{p_0, \delta}$.

For property (b), by Markov's inequality,
\[
P(V_{p, \delta}^*>c_{p, \delta}^*) \le \left( c_{p, \delta}^*\right) ^{-1} \mathsf{E} \left(\max_{t\in \mathbb{T}}  \frac{|\bar W_p(t)-2\bar{\Phi}(t)|}{[\bar \Phi (t)]^{\theta} }\right). 
\]
Let $B(t) = [\bar \Phi (t) ]^{-\theta} |\bar W_p(t)-2\bar{\Phi}(t)|$, and by the similar arguments as in Section \ref{sec:proof_discrete0.5}, we have
\[
\mathsf{E}[\max_{t\in \mathbb{T}} B(t)] \le C \sqrt{\log p} \cdot \max_{t\in \mathbb{T}} \mathsf{E} [B(t)].
\]
Further, $\mathsf{E} [B(t)] \le  [\bar \Phi (t) ]^{-\theta} (\mathsf{E}[\bar W_p(t)] + 2 \bar \Phi (t)) =  4 [\bar \Phi (t) ]^{1-\theta} \le 4$ for $\theta \in (1/2, 1]$. 

Summing up the above, we have
\[
P(V_{p, \delta}^*>c_{p, \delta}^*) \le C \left( c_{p, \delta}^*\right) ^{-1} \sqrt{\log p}< \alpha, 
\]
where the last step is by $c^*_{p, \delta} = C_0 \sqrt{\log p}$ with a large enough constant $C_0$.

Next, we demonstrate  the upper bound property of $\hat \pi_\delta^*$ with $\delta(t) = [\bar \Phi(t)]^{\theta}, \theta \in (1/2, 1]$. Similar arguments as in the proof of Theorem \ref{thm:discrete_0.5} for the upper bound can be applied with condition (\ref{cond:theta_05}) replaced by condition (\ref{cond:theta_1}). We omit the details to save space.

\bibliographystyle{chicago}
\bibliography{proportion_ref}

\end{document}